\pdfoutput=1

\documentclass[reprint, superscriptaddress, amsmath, amssymb, aps, prl, citeautoscript,longbibliography]{revtex4-2}

\usepackage{graphicx}
\usepackage{color}
\usepackage[svgnames]{xcolor}
\usepackage[colorlinks=True,linkcolor=DarkRed,citecolor=Blue,urlcolor=MediumBlue,pdfstartview=FitH,bookmarks=False,pdfpagemode=UseNone]{hyperref}

\begin{document}

\title{Terahertz metamaterials for light-driven magnetism}

\author{Matteo Pancaldi}
\email[]{matteo.pancaldi@unive.it}
\affiliation{Department of Molecular Sciences and Nanosystems, Ca' Foscari University of Venice, 30172 Venezia Mestre, Italy}

\author{Paolo Vavassori}
\affiliation{CIC nanoGUNE BRTA, 20018 Donostia-San Sebastián, Spain}
\affiliation{IKERBASQUE, Basque Foundation for Science, 48013 Bilbao, Spain}

\author{Stefano Bonetti} 
\email[]{stefano.bonetti@unive.it}
\affiliation{Department of Molecular Sciences and Nanosystems, Ca' Foscari University of Venice, 30172 Venezia Mestre, Italy}
\affiliation{Department of Physics, Stockholm University, 10691 Stockholm, Sweden}

\begin{abstract}
We describe the design of two types of metamaterials aimed at enhancing terahertz field pulses that can be used to control the magnetic state in condensed matter systems. The first structure is a so-called ``dragonfly'' antenna, able to realize a five-fold enhancement of the impinging terahertz magnetic field, while preserving its broadband features. For currently available state-of-the-art table top sources, this leads to peak magnetic fields exceeding 1 T. The second structure is an octopole antenna aimed at enhancing a circularly-polarized terahertz electric field, while preserving its polarization state. We obtain a five-fold enhancement of the electric field, hence expected to exceed the 1 MV/cm peak amplitude. Both our structures can be readily fabricated on top of virtually any material.
\end{abstract}

\maketitle

\section{Introduction}

The condensed matter physics community has recently shown an increased interest in the use of electromagnetic radiation in the terahertz range for the study of material systems \cite{salen2019matter}. The field of ultrafast magnetism was born almost three decades ago \cite{beaurepaire1996ultrafast}, but it is only in the last decade that intense terahertz radiation has been introduced in the game \cite{kampfrath2011coherent,kampfrath2013resonant,vicario2013off,bonetti2016thzdriven}. A major difference between more conventional near-infrared and terahertz pulses is that in the latter case the magnetic field component is slow enough to allow for a relatively strong Zeeman torque on the magnetization. In turn, this enables coherent control of the magnetization dynamics, in contrast to visible/near-infrared radiation where incoherent thermalization of non-equilibrium electron states is the dominant effect.

\begin{figure*}
    \includegraphics[width=2\columnwidth]{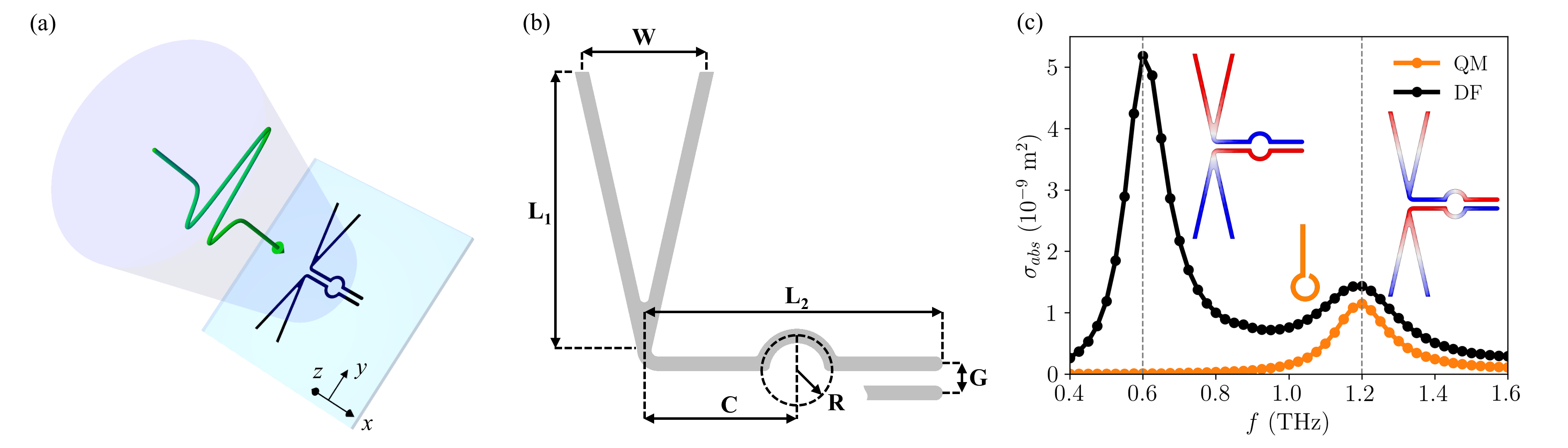}
    \caption{(a) Scheme of the DF antenna stimulated by a linearly-polarized (along the y-axis) quasi-single-cycle terahertz pulse. The $z=0$ plane represents the air/substrate interface. (b) Design of the DF antenna, for pointing out the relevant parameters: $W=30~\mathrm{\mu m}$, $L_1=L_2=60~\mathrm{\mu m}$, $C=30~\mathrm{\mu m}$, $R=7.5~\mathrm{\mu m}$, and $G=6~\mathrm{\mu m}$, with a track width of $3~\mathrm{\mu m}$. (c) Absorption cross sections calculated in the frequency domain for linearly polarized radiation at normal incidence, for both QM and DF designs. The insets show the normalized charge distribution (in the $\left[-1,1\right]$ range) of the oscillation modes corresponding to the absorption peaks for the DF antenna, and a scheme of the QM antenna, whose parameters were tuned to show a resonance at 1.2 THz. For that peak, the full width at half maximum for the (benchmark) QM antenna is 0.16 THz, while it is 0.21 THz for the DF antenna (calculated after removing a quadratic background).}
    \label{fig1}
\end{figure*}

Intense terahertz pulses can be obtained both at large-scale facilities \cite{back1999minimum,tudosa2004ultimate,oepts1995free-electron-laser,kovalev2018selective,dimitri2018coherent}, as well as in table-top setups \cite{hoffmann2011intense,fulop2020laser,sederberg2020tesla}. The latter are enabling the widespread use of single-cycle or few-cycles terahertz radiation to probe and control matter \cite{forst2011nonlinear,liu2012terahertz,hudl2019nonlinear,salen2019matter}. However, in order to fully explore dynamical regimes beyond the linear one, there is a need for terahertz pulses of even larger amplitude than what achievable at the source. To this end, much effort has been recently dedicated to metamaterial structures, i.e. patterned metallic thin-film antennas, aimed at a locally enhancing the linearly-polarized terahertz electric field \cite{razzari2011extremely,razzari2013terahertz,savoini2016thz,kang2018terahertz}. Surprisingly, only a few studies have been performed with the goal of enhancing the terahertz magnetic field component \cite{mukai2014antiferro,mukai2016nonlinear,polley2018thza,polley2018thzb}. Similarly, little attention has been paid towards designing terahertz metamaterials for enhancing fields with non-linear polarization.

In this article, we present the design, using finite element three-dimensional simulations, of two metamaterial structures aimed at filling these voids. For the first design, after revising the current approaches for enhancing the magnetic field component (mainly split-ring resonators), we present a ``dragonfly'' antenna which combines two different structures: a bow-tie broadband antenna \cite{compton1987bow-tie} and a coplanar stripline \cite{schnell2011nanofocusing}. The bow-tie antenna is the broadband receiving part, whereas the stripline transmits the signal in the region of interest, and it is responsible for the enhancement of the terahertz magnetic field. Such a design mostly preserve the broadband nature of the incoming terahertz field. For the second design, we build upon a strategy developed for radiation in the visible/near-infrared range \cite{biagioni2009cross,biagioni2009near-field}, and apply it to the terahertz range to obtain a metamaterial able to enhance circularly polarized electric fields \cite{gao2011terahertz}. Both proposed structures have dimensions in the tens of micrometer range, and can be readily fabricated using standard lithographic and deposition techniques.

\section{The dragonfly antenna}
\label{sec_df}

Current structures devoted to the enhancement of the magnetic field are based on the split-ring resonator geometry, where electromagnetic radiation at normal incidence can induce a current when the electric field is oriented perpendicularly to the gap \cite{katsarakis2004electric,garcia2005on,padilla2006dynamical}. In turn, the current generates a magnetic field normal to the plane of the split-ring, whose amplitude is enhanced with respect to the magnetic field component of the incoming radiation. Since the enhancement is local, i.e. confined inside the split-ring, such structures have to be patterned in close proximity to the magnetic feature of interest \cite{mukai2014antiferro,mukai2016nonlinear,qiu2018enhancing,zhang2023generation}. As a drawback, the split-ring design lacks flexibility, and its frequency response is characterized by narrow peaks in correspondence of the resonance modes. Through an equivalent circuit model, it is possible to recognize that the split-ring behaves as a LC circuit, and the magnetic resonance frequency is inversely proportional to the split-ring area \cite{zhou2005saturation,kafesaki2005left-handed}. The resonance tuning is then performed by varying the ring diameter, which also influences the extent of the field enhancement.

In order to decouple the resonance frequency from the field enhancement, a modified geometry can be considered, as shown by Polley et al. \cite{polley2018thza}. When a straight section is added to the split-ring, obtaining the so-called ``question mark'' (QM) geometry, the overall electrical length can be changed without altering the diameter of the ring. By doing so, the active region can maintain a constant size, independently from the target resonance frequency. However, such a structure still possesses narrow resonances, which can in principle be an unwanted feature when working with broadband pulses, and the impulsive characteristics of the excitation need to be preserved.

Here, we propose a design for increasing the bandwidth and preserve the frequency content of short terahertz pulses, by combining complementary geometries in the ``dragonfly'' (DF) design, as schematically shown in Fig. \ref{fig1}(a),(b). A bow-tie antenna \cite{compton1987bow-tie,runge2020spatial,rathje2023coupling} constitutes the broadband receiving part (characterized by the $W$ and $L_1$ dimensions), which is efficiently stimulated when aligned along the polarization axis of the impinging terahertz radiation (y-axis in the scheme). The electromagnetic field is then transmitted to the circular active region of radius $R$ via a coplanar stripline \cite{frankel1991terahertz,gupta1991subpicosecond,schnell2011nanofocusing} of length $L_2$ and pitch $G$. Our design aligns with the quest for metamaterials characterized by broadband electromagnetic functionalities, as reviewed in Ref. \cite{fan2019constructing}.

The electromagnetic properties of the DF antenna have been verified by means of finite element simulations in the frequency domain \cite{comsol}, with the dimensions specified in the caption of Fig. \ref{fig1}(b). A gold antenna, with a thickness of 100 nm, was placed on top of a semi-infinite crystalline quartz substrate. The material properties of gold were specified in terms of the electrical conductivity $\sigma=4.09 \cdot 10^7 ~\mathrm{S/m}$ \cite{wen2009dual}, while the substrate was characterized by a refractive index $n_{subs}=2.1$ (considered constant in the analyzed frequency range) \cite{davies2018temperature-dependent}. The other semi-infinite domain, where the incident wave comes from, is considered to be air, with a refractive index $n_{air}=1$. The dimensions have been chosen to work in a frequency range compatible with OH1, an organic crystal used for obtaining high-amplitude terahertz field in table-top setups via optical rectification \cite{brunner2008hydrogen-bonded}.

\begin{figure}
    \includegraphics[width=\columnwidth]{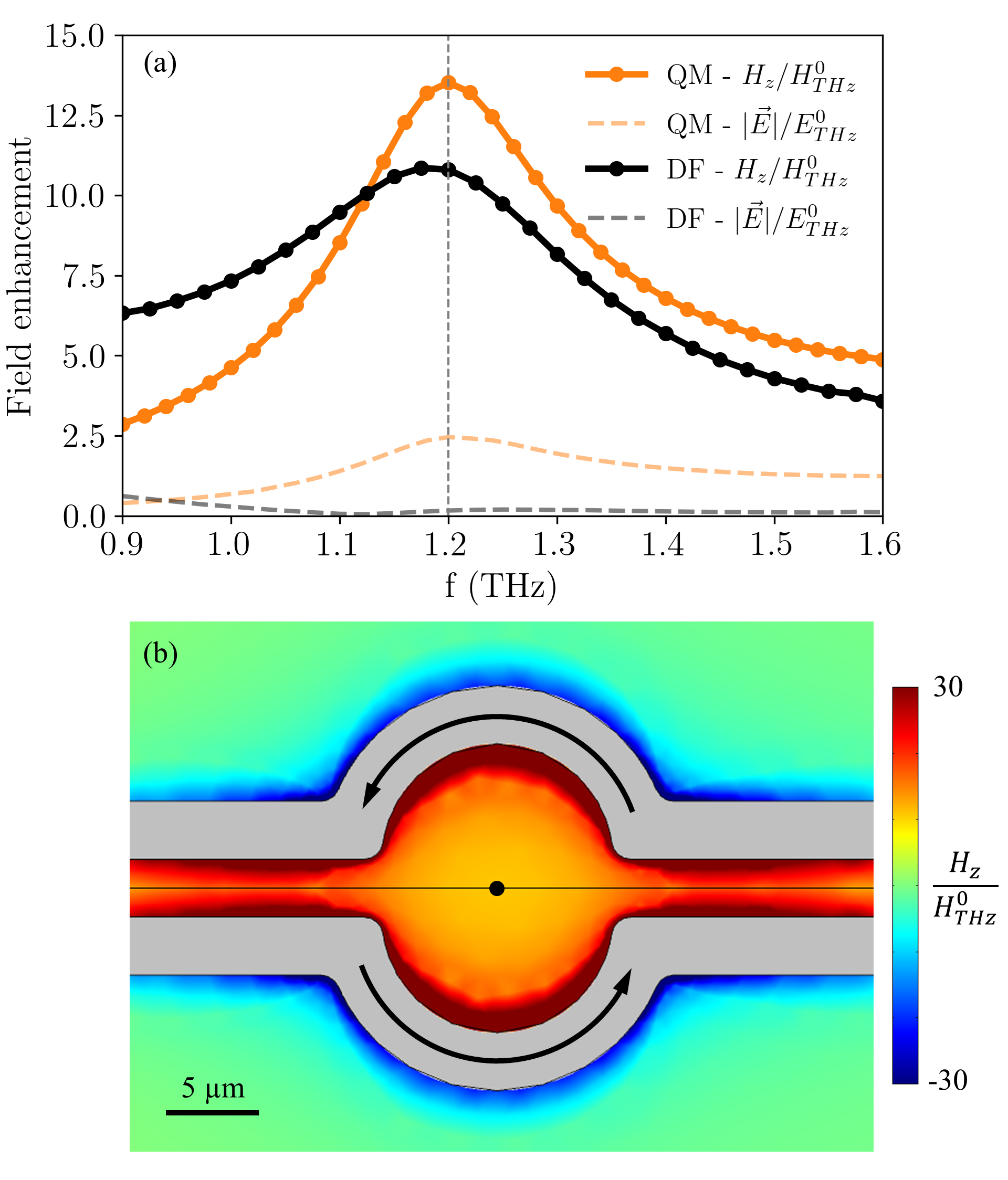}
    \caption{(a) Field enhancement at the center of the active region calculated in the frequency domain with respect to the incident field ($E_{THz}^0$, $H_{THz}^0$). Both the z-component of the magnetic field ($H_z$) and the magnitude of the electric field ($|\Vec{E}|$) are shown. (b) Map of the out-of-plane magnetic field enhancement for the dragonfly antenna at 1.2 THz in the $z=0$ plane. The arrows indicate the direction of the current in the coplanar stripline for generating a $H_z$ component in the positive z-axis direction.}
    \label{fig2}
\end{figure}

As a first characterization, Fig. \ref{fig1}(c) shows the absorption cross section, which is the ratio between the power absorbed by the antenna and the intensity of the incoming electromagnetic field. For the DF design, the absorption cross section shows two peaks in the considered frequency range. The charge distributions shown in the insets allow identifying the nature of the resonance modes: The first peak corresponds to the half-dipole resonance at 0.6 THz, while the second peak corresponds to the dipole resonance at 1.2 THz \cite{biagioni2012nanoantennas}. In both cases, the current flows in opposite directions in the two halves of the coplanar stripline, so determining an effective current loop in the circular active region. Besides being associated with a lower absorption, and hence with a lower average current, the active region lies close to the zeros of the surface charge for the 1.2 THz resonance, so a higher peak current can be obtained with respect to the 0.6 THz resonance. As a reference, Fig. \ref{fig1}(c) also shows the absorption cross section for the QM antenna, whose dimensions were selected in order to have the same resonance frequency and the same active region area. A comparison of the peak full width at half maximum for both geometries (0.16 THz for the QM, 0.21 THz for the DF) proves that the DF antenna possesses a larger bandwidth around the working frequency.

\begin{figure}
    \includegraphics[width=\columnwidth]{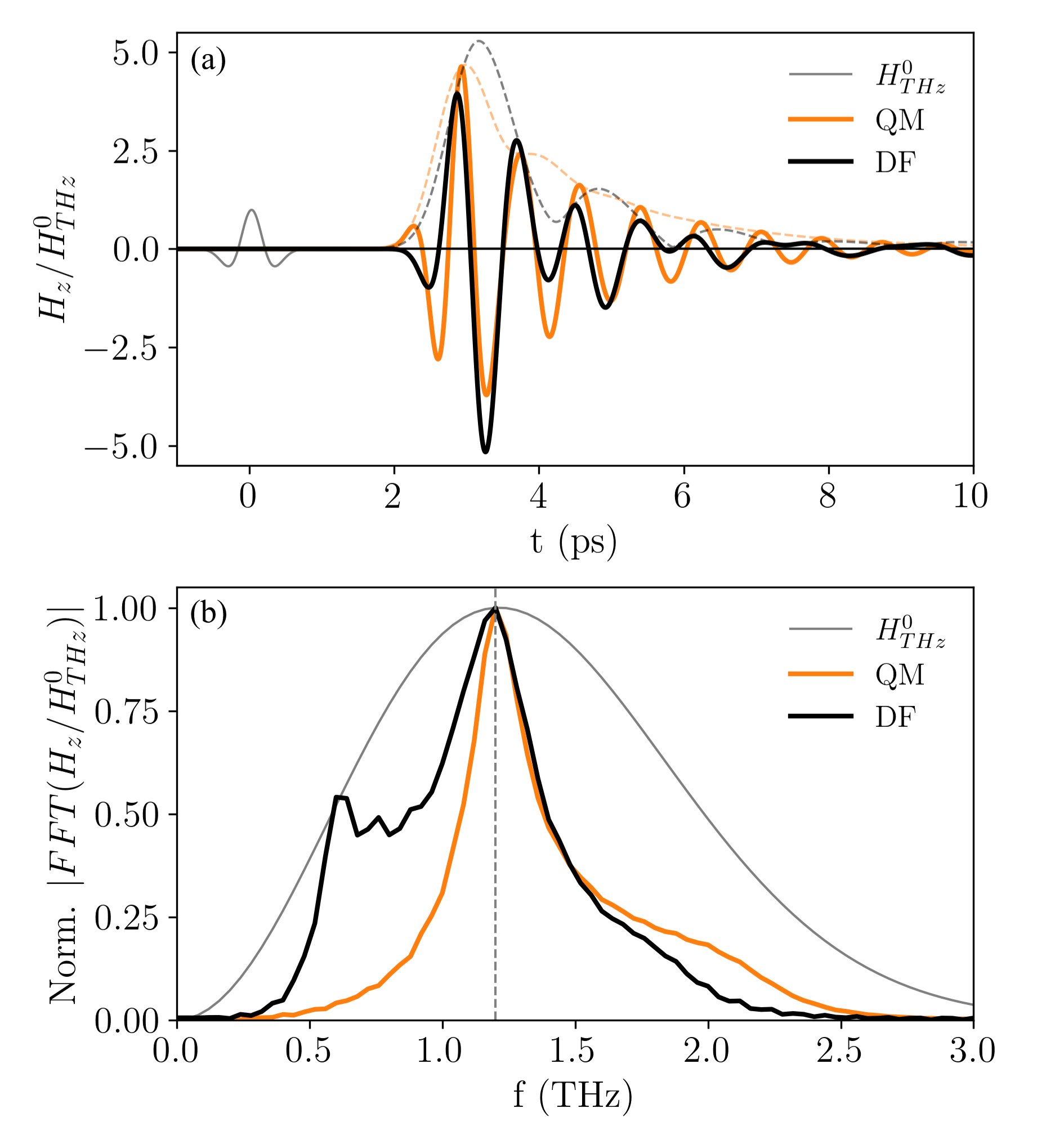}
    \caption{(a) Magnetic field enhancement at the center of the active region calculated in the time domain considering a quasi-single-cycle terahertz pump pulse (modeled as the second derivative of a Gaussian-shaped pulse). The dashed lines represent the signals' envelopes, indicating a faster decay for the DF geometry. (b) Normalized FFT amplitudes for the signals in panel (a). The wider bandwidth of the DF antenna helps in better preserving the temporal profile of the incident pulse.}
    \label{fig3}
\end{figure}

\begin{figure*}[t]
    \includegraphics[width=2\columnwidth]{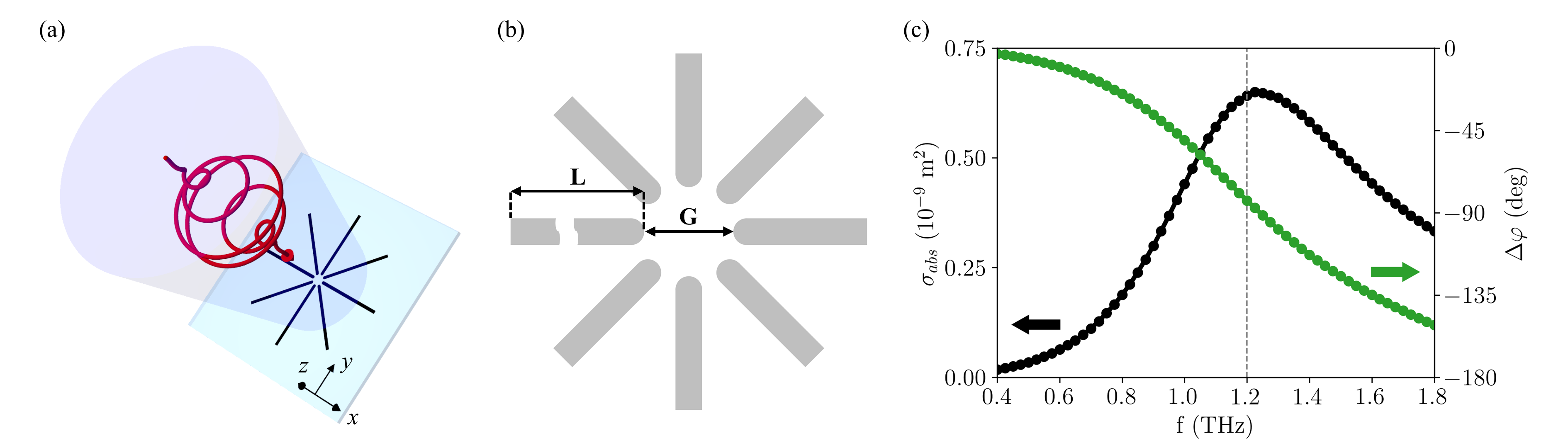}
    \caption{(a) Scheme of the octopole antenna stimulated by a circularly-polarized narrowband terahertz pulse. The $z=0$ plane represents the air/substrate interface. (b) Design of the octopole antenna, for pointing out the relevant parameters: $L=65~\mathrm{\mu m}$, and $G=10~\mathrm{\mu m}$, with a bar width of $3~\mathrm{\mu m}$. (c) Absorption cross section calculated in the frequency domain for circularly-polarized radiation at normal incidence. The plot also shows the relative phase, which is evaluated between the incident field and the field scattered by the antenna.}
    \label{fig4}
\end{figure*}

A comparison of the $H_z$ field enhancement at the center of the active region is reported in Fig. \ref{fig2}(a). The enhancement is calculated with respect to the incident electromagnetic field amplitude. At 1.2 THz, both the geometries show a comparable magnetic field enhancement (13.5 for the QM, 10.8 for the DF), but the DF antenna is associated to a flatter frequency response in the considered range, which is a consequence of the enhanced peak bandwidth and to the presence of the other resonance mode at 0.6 THz \cite{kumar2020bandwidth}. Moreover, due to its symmetry, the DF antenna is capable of enhancing the out-of-plane magnetic field while suppressing the electric field. This feature is relevant when it is important to decouple the two fields, which is not possible for a propagating electromagnetic wave, but it is feasible in the quasi-static regime \cite{jackson1999classical}. As shown by the black dashed line in Fig. \ref{fig2}(a), the electric field in the DF geometry is weakened with respect to the electric field component of the incident field. As a figure of merit, we introduce the ratio between the magnetic and electric field amplitudes, normalized to the impedance of free space ($Z_0$). For an electromagnetic wave travelling in vacuum (as the incident radiation in the considered simulations), we have that $Z_0H/E=1$. At 1.2 THz, we obtain that $Z_0H/E=65.3$ for the DF antenna, while $Z_0H/E=5.5$ for the QM antenna, showing that this latter geometry is less effective in shielding the active region from the incident electric field. The symmetry of the DF antenna is also relevant when considering the magnetic field enhancement map shown in Fig. \ref{fig2}(b). The enhancement grows uniformly from the center to the edge of the circular active region, and it is also present in the gap between the linear sections of the coplanar stripline (similarly to a pair of wires carrying opposite currents).

Finally, the effect of a broadband terahertz pulse has been simulated in the time domain, in order to validate all the results obtained in the frequency domain and to take into account a terahertz pulse with a realistic frequency spectrum \cite{brunner2008hydrogen-bonded}. As shown in Fig. \ref{fig3}(a), the incident quasi-single-cycle terahertz pulse was described considering the second derivative of a Gaussian-shaped pulse \cite{hoffmann2011intense}, with a standard deviation of 0.185 ps. Figure \ref{fig3}(a) also shows the $H_z$ enhancement as a function of time at the center of the active region: For both the DF and QM antennas, a five-fold enhancement can be obtained. Since the incident pulse has a broad spectrum, as reported in Fig. \ref{fig3}(b), the maximum enhancement over the whole incident pulse bandwidth is reduced with respect to the resonant enhancement shown in Fig. \ref{fig2}(a). However, for a state-of-the-art quasi-single-cycle terahertz magnetic field pulse with an amplitude of 300 mT, peak magnetic fields of the order of 1.5 T can be reached, which can be utilized (e.g.) for triggering non-linear magnetization dynamics in the few-ps time scale \cite{hudl2019nonlinear}. Moreover, by looking at the signals' envelopes highlighted by the dashed lines in Fig. \ref{fig3}(a), it is possible to notice that the oscillations following the main pulse (i.e. after 4 ps) have a faster decay for the DF antenna with respect to the QM. Those oscillations are a consequence of the filtering effect caused by the antenna bandwidth, which is less pronounced for the DF geometry, so confirming the increased bandwidth of this design. A comparison of the bandwidth for both designs can be obtained in the time domain by extracting the amplitude of the Fourier transform, as shown in Fig. \ref{fig3}(b). Again, the wider resonance peak at 1.2 THz and the presence of the 0.6 THz peak make the DF design better suited for preserving the impulsive characteristics of the pump pulse.

\section{The octopole antenna}
\label{sec_octo}

Another situation of recent growing interest for the study of light-driven magnetism is the one of enhancing the terahertz electric field while preserving the circular polarization of the excitation. Historically, circularly-polarized light has often been used to investigate magnetism in materials. Most prominent is the case of the magnetic circular dichroism at x-ray or UV wavelengths, which exploits the different absorption coefficient of light with opposite helicity to quantitatively retrieve the spin and orbital contribution of the overall magnetic moment of a sample \cite{stohr2006magnetism}. This effect is present and has been used even in the optical regime, although quantitative material properties are not easily extracted in this case. Going further down in the photon energy, reaching the terahertz range, circularly-polarized light has been recently employed not to probe but rather to drive novel magnetic states in matter. As representative example, we cite the dynamical multiferroicity effect in diamagnetic insulators \cite{juraschek2017dynamical}, which has been recently shown to be enhanced by the Barnett effect in two different experiments, leading to large magnetic moments driven by circularly-polarized terahertz fields \cite{basini2022terahertz,davies2023phononic}. It seems therefore timely to design metamaterials capable of enhancing such fields, which are anticipated to allow for enhanced control of magnetic states in matter.

The design that we present for this case is a variation of the cross antenna proposed in Refs. \cite{biagioni2009cross,gao2011terahertz}, where we add two additional pairs of antennas along the diagonals, see the schematics in Fig. \ref{fig4}(a). We name this geometry the ``octopole'' antenna, for obvious reasons. Each of the electrodes of the antennas, of length $L$, is placed at an angle of 45 degrees with respect to the nearest neighboring one, creating a gap $G$ between opposite electrodes, as shown in Fig. \ref{fig4}(b). Through finite element simulations \cite{comsol}, a 100 nm-thick gold octopole antenna was designed to resonate at a frequency of 1.2 THz, as indicated by the absorption cross section in Fig. \ref{fig4}(c). The figure shows a peak of the absorption at the design frequency and, consistently, a 90 degrees phase difference $\Delta\varphi$ of the scattered radiation as compared to the incident field. All the material parameters considered in the simulations are the same as the ones presented in the previous section for the DF antenna.

\begin{figure}
    \includegraphics[width=\columnwidth]{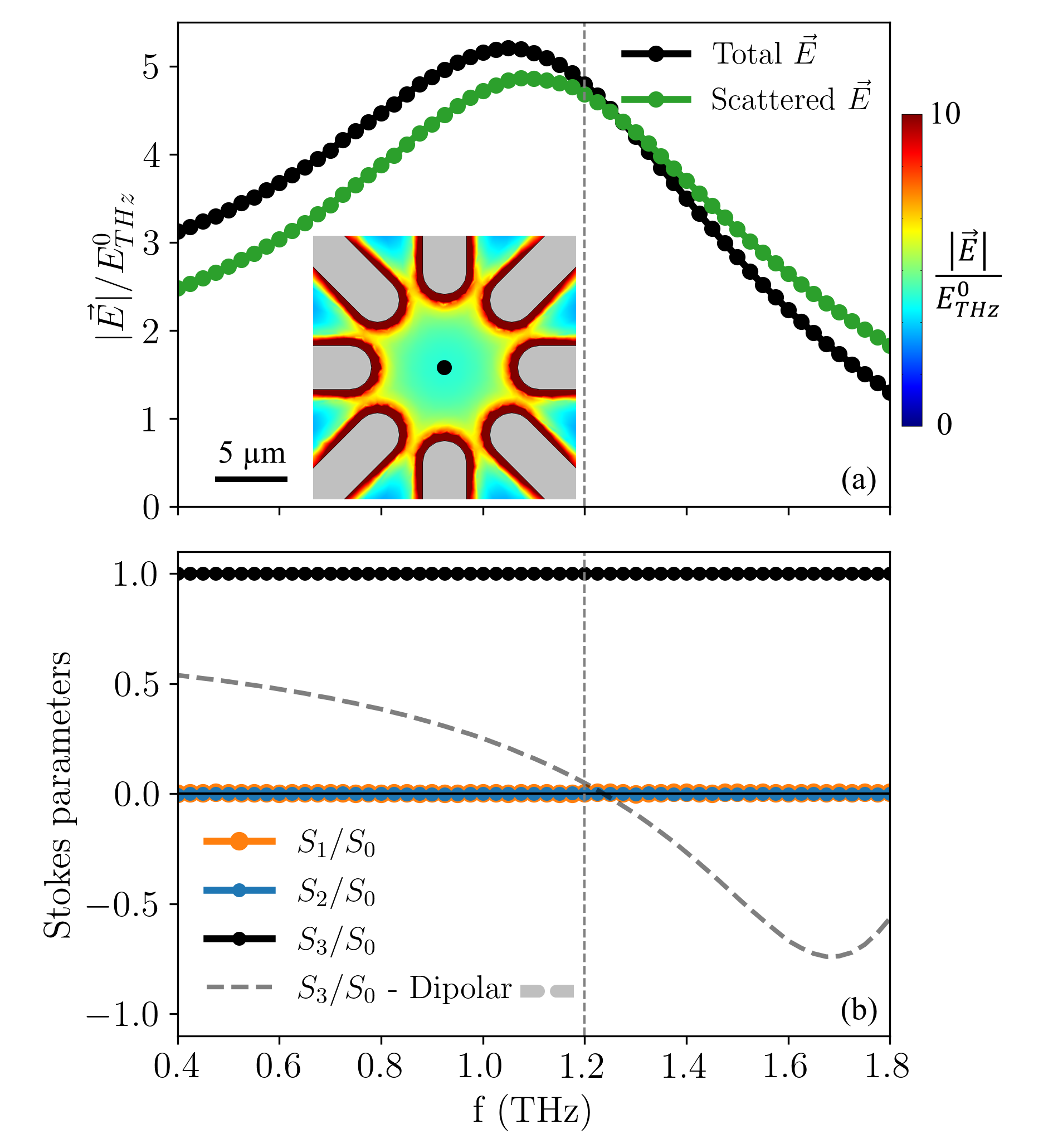}
    \caption{(a) Electric field enhancement at the center of the active region calculated in the frequency domain with respect to the incident field $E_{THz}^0$. The black curve represents the total electric field, while the green curve only considers the electric field scattered by the metamaterial. The inset shows a map of the electric field enhancement for the $z=0$ plane at 1.2 THz in the central part of the metamaterial. (b) Normalized Stokes parameters as a function of frequency, for quantifying the polarization state in the center of the active region. As a comparison, the dashed line shows the $S_3/S_0$ parameter for a dipole antenna.}
    \label{fig5}
\end{figure}

Figure \ref{fig5}(a) shows the electric field enhancement (both the total one, in black, and the scattered one, in green) in the center of the octopole antenna. The simulated data indicates an approximate five-fold enhancement in the frequency range between 1 and 1.2 THz. For a state-of-the-art narrowband terahertz electric field pulse of 200 kV/cm amplitude, this would mean that MV/cm electric fields are achievable. The two-dimensional map in the inset of Fig. \ref{fig5}(a) reveals that the enhancement at 1.2 THz is rather uniform in a central circular area of diameter $G/2$, and the uniformity is improved by rounding the end of the bars composing the metamaterial. In order to characterize the degree of polarization in the octopole antenna, we calculated the Stokes parameters $S_1$, $S_2$ and $S_3$, and normalize them to $S_0$. From the Stokes parameters, a full analysis of the polarization state can be obtained \cite{hulst1981light}: While the $S_0$ parameter represents the total intensity, $S_1$ and $S_2$ are associated with linear polarization along two sets of orthogonal axes, and $S_3$ is associated with circular polarization. The results are shown in Fig. \ref{fig5}(b). The simulations show a perfect degree of circular polarization at the center of the metamaterial, with the $S_3/S_0$ ratio being equal to 1 in the range from 0.4 to 1.8 THz, where $S_1/S_0$ and $S_2/S_0$ are negligible. Such a behavior can be compared with the $S_3/S_0$ ratio calculated for a dipole antenna \cite{razzari2011extremely,savoini2016thz} with the same $L$ and $G$ dimensions used for the octopole. In such a case, the circular polarization is totally lost at the resonance frequency, and it shows a non-monotonous trend over the same range.

\section{Conclusion}

We designed two novel terahertz metamaterials aimed at filling up the void left by conventional structures of this kind. In the first design, we proposed a ``dragonfly'' structure able to enhance the terahertz magnetic field of a broadband, single-cycle pulse in the out-of-plane direction. Our design greatly extends the overall bandwidth of enhancement, resulting in a five-fold peak enhancement in the time domain, uniform over an area of several micrometers in lateral size. This would allow for quasi-single-cycle local magnetic field pulses with amplitude of more than 1 T. We anticipate that such intense and short magnetic fields will open up for the study of nonlinear magnetization dynamics in several magnetic systems, including antiferromagnets and altermagnets \cite{smejkal2022emerging}. In the second design, we suggested an octopole antenna design able to reach a five-fold enhancement of an incident narrowband, circularly-polarized terahertz electric field, retaining its polarization state, uniformly over an area with lateral size of several micrometers in this case as well. Our design allows for reaching the MV/cm electric field amplitude regime, a regime where nonlinear effects become evident. For the case of insulating materials, the implementation of our structures in experiments is particularly straightforward, as they can be directly patterned on top of the materials.

M.P. and S.B. acknowledge support from the Italian Ministry of University and Research, PRIN2020 funding program, Grant No. 2020PY8KTC. S.B. acknowledges support from the Knut and Alice Wallenberg Foundation, Grants No. 2017.0158 and 2019.0068. P.V. acknowledges support from the Spanish Ministry of Science and Innovation under the Maria de Maeztu Units of Excellence Programme CEX2020-001038-M, and the project PID2021-123943NB-I00 (MICINN/FEDER).

\bibliography{dragonfly_bib.bib}

\end{document}